\documentclass[pra,showpacs,preprintnumbers,amsmath,amssymb,tightenlines,twocolumn]{revtex4}
\usepackage{graphicx}
\usepackage{dcolumn}
\usepackage{bm}
\usepackage{epsfig}
\usepackage{stmaryrd}
\usepackage{tensor}
\usepackage{color}  

\newcommand{\ii}{\text{i}}

\newcommand{\cc}{\text{c}}
\newcommand{\mm}{\text{m}}

\newcommand{\bra}[2][]{\mathinner{\langle #2\rvert}_{#1}}
\newcommand{\ket}[2][]{\mathinner{\lvert#2\rangle}_{\hspace{-0.1em}#1}}

\newcommand{\sub}[2]{{#1}_{\mbox{\!\! \scriptsize #2}}}

\newcommand{\bv}[1]{\mathbf{ #1 }}

\def\beq{\begin{equation}}
\def\eeq{\end{equation}}

\def\CR{\nonumber\\[0.15cm]}

\newcommand{\rref}[1]{Ref.~\cite{#1}}
\newcommand{\fref}[1]{Fig.~\ref{#1}}
\newcommand{\frefp}[2]{Fig.~\ref{#1}~(#2)}

\newcommand{\eref}[1]{Eq.~(\ref{#1})}

\newcommand{\sref}[1]{section~\ref{#1}}

\newcommand{\cref}[1]{chapter~\ref{#1}}

\newcommand{\Cref}[1]{Chapter~\ref{#1}}

\newcommand{\aref}[1]{appendix~\ref{#1}}
\newcommand{\bref}[1]{(\ref{#1})}

\definecolor{d_green}{RGB}{0,128,0}

\binoppenalty=\maxdimen
\relpenalty=\maxdimen

\begin{document}

\title{Coupling of a nano mechanical oscillator and an atomic three-level medium}
\author{A.~{Sanz-Mora}}
\affiliation{Max Planck Institute for the Physics of Complex Systems, N\"othnitzer Strasse 38, 01187 Dresden, Germany}
\author{A.~Eisfeld}
\affiliation{Max Planck Institute for the Physics of Complex Systems, N\"othnitzer Strasse 38, 01187 Dresden, Germany}
\author{S.~W\"uster}
\affiliation{Max Planck Institute for the Physics of Complex Systems, N\"othnitzer Strasse 38, 01187 Dresden, Germany}
\author{J.-M.~Rost}
\affiliation{Max Planck Institute for the Physics of Complex Systems, N\"othnitzer Strasse 38, 01187 Dresden, Germany}
\begin{abstract}
We theoretically investigate the coupling of an ultracold three-level atomic gas and a nano-mechanical mirror via classical electromagnetic radiation. The radiation pressure on the mirror is modulated by absorption of a probe light field, caused by the atoms which are electromagnetically rendered nearly transparent, allowing the gas to affect the mirror. 
In turn, the mirror can affect the gas as its vibrations generate opto-mechanical sidebands in the control field. We show that the sidebands cause modulations of the probe intensity at the mirror frequency, which can be enhanced near atomic resonances. Through the radiation pressure from the probe beam onto the mirror, this results in resonant driving of the mirror. Controllable by the two photon detuning, the phase relation of the driving to the mirror motion 
decides upon amplification or damping of mirror vibrations. 
This permits direct phase locking of laser amplitude modulations to the motion of a nano-mechanical element opening a perspective for cavity-free cooling through coupling to an atomic gas.
\end{abstract}
 
\maketitle

\section{Introduction}
%
The manipulation of an ever more diverse variety of nano-mechanical oscillators \cite{poot:mechquantumsyst} using the intricate control over electromagnetic fields provided by quantum optics is the subject of quantum opto-mechanics \cite{aspelmeyer:review,kippenberg:review}. Interfacing light-fields in tailored quantum states with mechanical systems deeply in the quantum regime promises applications in quantum information transfer between different spectral realms \cite{Bochmann_Cleland:microwave_optical}, studies of the quantum-classical transition \cite{buchmann:mediated_int} as well as new impulses for fundamental physics \cite{pikovski:planckquantop}, predominantly gravitational wave detection \cite{mccleland:advanced_interferometry,Sawadsky_2015_of}.

A key benefit of nano-mechanical systems is their coupling to electromagnetic radiation over a wide range of the spectrum. This facilitates interfacing with diverse quantum devices, such as optical cavities \cite{aspelmeyer:review}, Josephson circuits \cite{Pirkkalainen:josephson_optomech} or quantum dots \cite{Montinaro:quant_dot_nanowire} in hybrid setups. A newly emerging group of hybrid setups involves atomic or molecular systems \cite{camerer:latticecloud,hammerer:latticecloudtheory,Genes_2011_cooling,bariani:controlwsingleatom,hammerer:EPR,singh:wignerntomog,singh:cantilevermol,bariani:hybridlambda,genes:cooling_gain,dantan:hybrid_doped,vogell:longdistcouple,zhang:lambdacooling}. They enable the exploitation of the versatile toolkit of cold atom quantum manipulations for the control of mechanical systems. Recent work has established that coupling internal states of atomic or molecular ensembles to nano-mechanical oscillators yields intriguing features, such as atom-mirror entanglement \cite{hammerer:EPR,singh:wignerntomog} and mechanical squeezing \cite{singh:cantilevermol}. 

\begin{figure}[htb]
\centering
\includegraphics[trim=0cm 3cm 0cm 3cm, width=0.99\columnwidth]{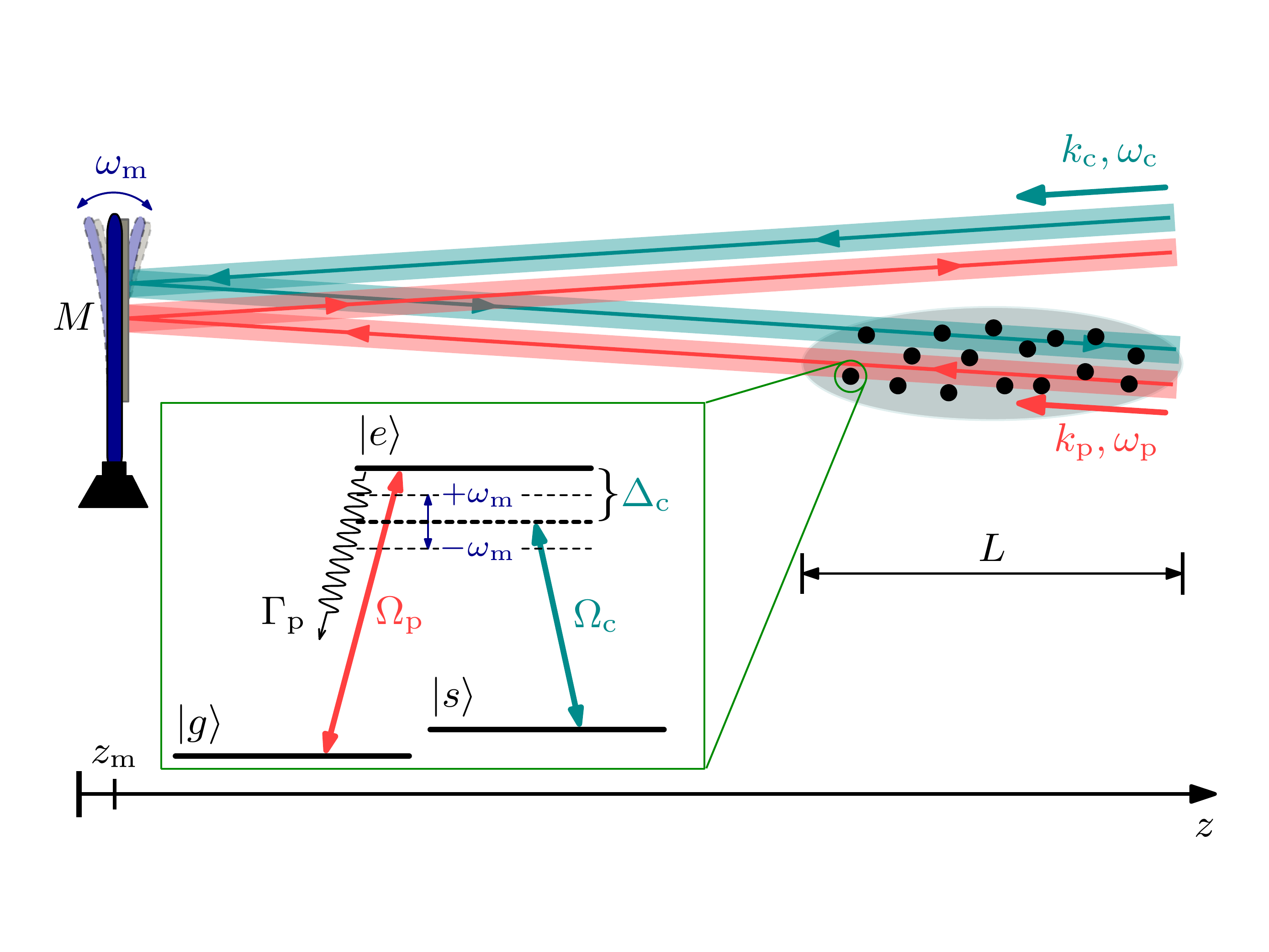}
\caption{(color online) Schematic diagram of EIT medium (dots) coupled to mechanically oscillating mirror via probe and control lasers with different optical paths. (inset) Energy level diagram of the EIT medium, realizing a  $\Lambda$ scheme. We also indicate the control laser sidebands due to mirror vibrations and spontaneous decay \cite{footnote:decay}. 
\label{setup}}
\end{figure}

Here we present a novel scheme to affect nano-mechanical oscillators in the classical regime that does not require a cavity, in contrast to many of the  proposals listed above. Instead, control of the mechanical motion of a mirror is achieved by coupling it to an ultra-cold gas with running wave laser fields \cite{camerer:latticecloud,hammerer:latticecloudtheory,vogell:longdistcouple}.
 In our case, the atoms from the cold gas interact with two laser beams under the condition of electro-magnetically induced transparency (EIT) \cite{fleischhauer:review}, as sketched in \fref{setup}. An EIT control beam is reflected by the mirror before interacting with the atomic gas. Any vibrations of the mirror imprint a phase modulation onto this EIT control beam, producing side-bands of the control field detuned by the mirror frequency. This causes a modulation of the intensity of the transmitted probe beam with the mirror frequency. This effect is maximal when the mirror frequency matches the energy gap between two atomic eigenstates. The probe beam causes driving of the mirror at its resonance frequency through radiation pressure. Whether this driving amplifies or damps the mirror motion depends on the relative phase shift between probe beam amplitude modulations and mirror oscillation.

We show that this relative phase shift can be adjusted by choice of the overall two-photon detuning of the EIT lasers. At the semi-classical level discussed here, the scheme allows phase-locking the amplitude modulations of a laser to motion of a mechanical element. Equivalently, the atomic cloud allows the conversion of phase-modulations of one light-field (the control beam), into amplitude modulations of another (the probe beam).

This article is organized as follows: In \sref{setupsection} we discuss our setup of mirror, atomic cloud and light fields followed by the physical model describing this arrangement in \sref{model}. Subsequently we analyse the dynamical response of the system, first of the atomic medium to a constantly oscillating mirror, \sref{periodic}, and then of the mirror being driven by the response of that medium, \sref{fullycoupled}.
In \sref{applicability} we investigate in which parameter regime the ensuing coupling between mirror and medium shows prospects for manipulations of the mirror, before concluding in \sref{conclusion}.

\section{Setup}
\label{setupsection}

%
Our atom-optomechanical setup consists of an ensemble of trapped, non-interacting ultra cold atoms, coupled to a mirror of mass $M$, see \fref{setup}. The  centre of mass position  $z_{\text{m}}$ of the mirror may oscillate around its equilibrium position $z=0$ with frequency $\omega_{\text{m}}$. For the atoms, we consider three relevant internal electronic states, $\ket{g}$, $\ket{s}$ and $\ket{e}$. The states $\ket{g}$, $\ket{s}$ are long-lived meta-stable ground states, while $\ket{e}$ decays to $\ket{g}$ with a rate $\Gamma_{\text{p}}$, as sketched in the inset of \fref{setup}.  Each of two laser beams pass through the atom cloud and reflects once from the mirror. The probe beam (wavenumber $k_{\text{p}}$, frequency $\omega_{\text{p}}$) couples the states $\ket{g}$ and $\ket{e}$ resonantly with Rabi frequency $\Omega_{\text{p}}$. It passes through the atomic cloud before reflecting off the mirror and leaving the system. The control beam (wavenumber $k_{\text{c}}$, frequency $\omega_{\text{c}}$) couples the states $\ket{s}$ and $\ket{e}$ with Rabi frequency $\Omega_{\text{c}}$ and detuning $\Delta_{\text{c}}$. In contrast to the probe beam, it reflects off the mirror first, then passes through the cloud and finally leaves the system. 

A central feature of our setup is that the two light beams are operated under typical conditions for EIT, $\Omega_{\text{p}} \ll \Omega_{\text{c}}$. At the EIT resonance $\Delta_{\text{c}} = 0$, atoms in the medium settle into a so called dark state, $\ket{\text{d\,}} \sim \Omega_{\text{c}} \ket{g} - \Omega_{\text{p}} \ket{s}$, in which excitation to the decaying state $\ket{e}$ is suppressed through quantum interference, causing the gas to become transparent for the probe beam~\cite{fleischhauer:review}. Since this transparency is a subtle quantum interference phenomenon it constitutes a sensitive probe allows sensitive probing of the coupling to the mechanical oscillator, which perturbs the EIT conditions and therefore is expected to have a noticeable effect.

Since the control beam is reflected off the vibrating mirror surface, the time dependent boundary condition on its electromagnetic field causes a modulated Rabi-frequency  
\begin{align}
\Omega_{\text{c}}(t) &= \Omega_{\text{c}} \text{exp}[\ii\,k_{\text{c}}z_{\text{m}}(t)] \approx \Omega_{\text{c}} [1 + \ii\,k_{\text{c}}z_{\text{m}}(t)],
\label{omegacmodul}
\end{align}
which will provide the desired perturbation of perfect EIT conditions.  In the last step of \bref{omegacmodul} we assume that the mirror displacement is small compared to the optical wavelength, although this simplification is not crucial for the physics described later. For constant harmonic motion of the mirror, $z_{\text{m}}(t)=z_0 \cos{(\omega_{\text{m}} t)}$, the power spectrum of the control Rabi frequency acquires sidebands $\omega_{\text{c}}\,\pm\,\omega_{\text{m}}$ as in multi-chromatic EIT \cite{Wanare:multicolCPT,Li:threelevelbichrom,LiJun:threelevelbichrom,HongJu:trichromkerr,zhang:polychromtransp}. We will show that the phase modulation of the control fields causes a time-dependent modulation of the transmission of the probe beam through the medium, 
or in short, the phase modulation of the control beam is turned into an amplitude modulation of the probe beam.

Due to the radiation pressure exerted by the probe beam on the mirror, we obtain a closed feedback loop, where the running wave fields are used to separately mediate the two directions of 
mutual coupling between the nano-mechanical mirror and the EIT medium. 

\section{Model}
\label{model}

We now formalise the setup presented in the preceding section, treating the light fields and the mirror classically, but the atomic EIT medium quantum-mechanically. This is valid for sufficiently large amplitudes of mirror motion compared to the zero-point motion, and optical fields that are sufficiently coherent and intense to neglect quantum fluctuations.

\subsection{Mirror}
\label{mirror}

The classical mirror is described by Newton's equation for a driven harmonic oscillator
\begin{align}
M \ddot{z}_{\text{m}}(t)+M \tensor*{\omega}{^{2}_{\text{m}}}\tensor*{z}{_{\text{m}}}(t) &= F(t) \text{,}
 \label{newton}
\end{align}
where $F(t)$ is the external driving force due to the radiation pressure by the probe and control beams given by 
\begin{align}
F(t)&=2 [W_{\text{p}}(t) +  W_{\text{c}}]/c\,.
 \label{radpressure}
\end{align}
The power $W_{\text{p}}(t)$ of the probe beam reflecting off the mirror may be time dependent due to varying transmission properties of the atomic medium. In contrast the reflected control beam power $W_{\text{c}}$ is constant as the beam only passes the medium that could absorb it \emph{after} reflection off the mirror.

Under conditions of perfect EIT, that is $\Delta_{\text{c}}=0$ and without modulations of the coupling beam, the reflected probe beam power would be the incoming probe beam power $W_{\text{p}}(t)=W_{\text{p}0}$. However, since the control beam modulates the transmission properties of the atomic medium, the probe beam power impinging on the mirror will be a function of the incoming probe beam power and time, i.e.~$W_{\text{p}}(t)=f(W_{\text{p} 0},t)$. To determine the function $f$, we have to study the atomic medium, which is described in \sref{atomicmedium}.

The model could easily be extended to include intrinsic damping and driving of the mirror induced by its coupling to a thermal environment at a finite temperature due to the mirror clamping. 

\subsection{Atomic medium}
\label{atomicmedium}

The atomic medium consists of $N$ non-interacting atoms at positions $\bv{r}_n$. The interaction of each atom with the two laser beams is described in the dipole- and rotating wave approximation by the internal Hamiltonian
\begin{align}
\hat{H}^{(n)}/\hbar&=\frac{1}{2}(\Omega_{\text{c}}(\bv{r}_n,t) \hat{\sigma}_{es}^{(n)} - \Omega_{\text{p}}(\bv{r}_n,t) \hat{\sigma}_{eg}^{(n)} + \text{h.~c.}) 
\CR
&+\Delta_{\text{c}} \hat{\sigma}_{ss}^{(n)}, 
\label{singleatomhamil}
\end{align}
where transition operators $\hat{\sigma}_{\beta \alpha}^{(n)}=[\ket{\beta}\bra{\alpha}]_{n}$ act on atom $n$ only.

The density matrix for the $n$'th atom, $\hat{\rho}^{(n)}$ evolves according to a Lindblad master equation
\begin{align}
\dot{\hat{\rho}}^{(n)}=\frac{\ii}{\hbar} [\hat{H}^{(n)},\hat{\rho}^{(n)}] + {\cal L}[\hat{\rho}^{(n)}] \text{,}
\label{mastereqn}
\end{align}
where the super-operator $ {\cal L}$ describes spontaneous decay of atom $n$ from level $\ket{e}$ to $\ket{g}$ \cite{drake:atomicphysics,footnote:decay}, and thus $ {\cal L}[\hat{\rho}^{(n)}]=\hat{L}_n\hat{\rho}^{(n)}\hat{L}_n^\dagger - (\hat{L}_n^\dagger\hat{L}_n\hat{\rho}^{(n)}+\hat{\rho}^{(n)}\hat{L}_n^\dagger\hat{L}_n)/2$ with decay operator $\hat{L}_n=\sqrt{\Gamma_{\text{p}}}\hat{\sigma}_{\beta \alpha}^{(n)}$.

Since the light fields causing the couplings $\Omega_{\text{p,c}}(\bv{r}_n)$ in \eref{singleatomhamil} are affected by the response of the atoms in the medium through which they propagate, \eref{mastereqn} has to be solved jointly with the optical propagation equations for the light fields (Maxwell-Bloch equations).   
However, it is known that for c.w.-fields, the medium settles into a steady state beyond some initial transient time, providing an optical susceptibility $\chi(\bv{r})=2d_{eg}^2\rho_{ge}(\bv{r})/[\hbar \epsilon_0 \Omega_{\text{p}}(\bv{r})]$ for the probe beam, where $d_{eg}$ is the transition dipole moment of the probe transition and $\rho_{ge}(\bv{r}) = \sum_n \langle \hat{\sigma}_{ge}^{(n)} \rangle \delta(\bv{r} - \bv{r}_n)$ the collective atomic coherence.
In the linear regime and for a homogeneous complex susceptibility $\chi(\bv{r})\equiv \chi' + \ii \chi''$ the transmitted power through a medium of length $L$ is $W = W_0 \exp{[- k_{\text{p}} L \chi'']}$, where $W_0$ is the incoming power (we split complex numbers as $z=z'+ \ii z''$ into real part $z'$ and imaginary part $z''$). 

For the setup in \fref{setup} the phase modulation \bref{omegacmodul} of the control Rabi frequency precludes a genuine steady state. However, if the modulation period is slow enough compared to the time it takes probe beam phase-fronts to pass through the medium, we can nonetheless obtain a simple response of the medium, as argued in \aref{optics}. The medium is then described by a time-dependent susceptibility $\chi(t) = S \rho_{ge}(t)$ with $S=2d_{eg}^2{\cal N}/[\hbar \epsilon_0 \Omega_{\text{p}}]$. Here, $\rho_{ge}(t)$ is determined from the solution of \eref{mastereqn} for a single atom standing representative for the entire medium, and ${\cal N} = \sum_n \delta(\bv{r} - \bv{r}_n)$ is the density of the medium. For this solution of \bref{mastereqn} including coupling to the mirror, we assume the following probe power to impinge on the mirror:
\begin{align}
W_{\text{p}}(t)&=W_{\text{p}0}\exp{\left[- k_{\text{p}} L  \chi''(t) \right]}
\CR
&\approx W_{\text{p}0}(1- A \rho_{ge}''(t)),
\label{probetransmisison}
\end{align}
with $ A=k_{\text{p}} L S = d \: \Gamma_{\text{p}}/\Omega_{\text{p}}$,
where we have used the optical depth $d=6 \pi {\cal N} L k_{\text{p}}^{-2}$ of the medium. This specifies the function $f$ of \sref{mirror}.

Using the power \bref{probetransmisison} for the probe radiation pressure \bref{radpressure} and \eref{omegacmodul} for the phase modulation of the control beam, the master equation \bref{mastereqn} and Newton's equation \bref{newton} become a coupled system of differential equations.

\subsection{Light fields}
\label{light}

The semi-classical model of the preceding two sections treats the propagating probe and control beams as classical electro-magnetic fields. It further neglects the travel time of optical beams between mirror and all atomic positions in the atom cloud, which hence has to be much shorter than the dynamical time scale of the problem that we study. The latter time-scale is given by the mirror period $T_{\text{m}}=2\pi/\omega_{\text{m}}$, so that the above assumptions are well satisfied for mirrors with frequencies in the MHz-GHz range and typical optical path lengths.

\section{Vibrating mirror coupled to atomic cloud}

In the following we analyse the consequences of coupling a vibrating mirror to an atomic $\Lambda$-type EIT medium with the model developed in \sref{model}. In a first step, we take into account the phase-modulation of the control beam by the vibrating mirror, but neglect all radiation pressure on the mirror. This yields an analytically solvable time-periodic model, presented in \sref{periodic}. In a second step, we close the feedback loop by incorporating radiation pressure on the mirror. As shown in \sref{fullycoupled} this gives rise to interesting dynamics, which can be understood using the results of \sref{periodic}.

\subsection{Time-periodic model}
\label{periodic}

If the driving force $F(t)$ is neglected in \eref{newton}, the mirror will undergo harmonic oscillations $z_{\text{m}}(t)=z_0 \cos{(\omega_{\text{m}} t)}$ with amplitude $z_0$. These oscillations give rise to constant strength sidebands in the control light field $ \Omega_{\text{c}}(t)=\Omega_{\text{c}}[ 1  +  \eta (\text{e}^{\ii \omega_{\text{m}} t} + \text{e}^{-\ii \omega_{\text{m}} t})/2]$, with relative amplitude $\eta = k_{\text{c}} z_0$. This prevents the atomic system \bref{mastereqn} from settling into a genuine steady state, which suggests the construction of an asymptotic solution in terms of Fourier components of the density operator: $\hat{\rho}=\sum_{k=-\infty}^{\infty} \hat{\rho}_k \exp{[-\ii k \:\omega_{\text{m}} t]}$, see for example \rref{Wanare:multicolCPT}. For long times ($\Gamma_{\text{p}}t \gg 1$) we demand the Fourier amplitudes to become steady 
\begin{align}
\frac{\partial}{\partial t} \hat{\rho}_k = 0\,.
\label{steadyfourier}
\end{align}
Due to the presence of sidebands, \eref{steadyfourier} and \eref{mastereqn} create an infinite hierarchy of coupled equations for the $\hat{\rho}_k$. We truncate the hierarchy at second order by neglecting all $\hat{\rho}_k$ with $|k|>1$, in what  amounts to a first order perturbative expansion in $\eta$. 
We thus keep only a constant density operator $\hat{\rho}_0$ and its first harmonics at the mirror frequency $\hat{\rho}_{\pm}$, 
\begin{align}
\hat{\rho}(t) \simeq \hat{\rho}_{0} + \hat{\rho}_{+}\text{e}^{-\ii\omega_{\text{m}}t} +\hat{\rho}_{-}\text{e}^{\ii\omega_{\text{m}}t}.
\label{rhoexpand}
\end{align}
We are now interested in modulations of the imaginary part of the probe coherence $\rho_{ge}'' = \text{Im}[\rho_{ge}]$ (as before we split complex numbers as $z=z'+ \ii z''$ into real part $z'$ and imaginary part $z''$). These modulations will 
affect  absorption by the medium according to \eref{probetransmisison}.
We define 
\begin{align}
\rho_{ge}''(t) = {\rho}_{0,ge}''  + {\delta\rho}_{ge}'' \cos{(\omega_{\text{m}} t + \alpha)},
\label{coherencemodul}
\end{align}
where now $\alpha$ is the relative phase between absorption modulations and mirror motion and $ {\delta\rho}_{ge}'' $ is the (real) amplitude of such modulations.

From our solution of \eref{steadyfourier} we find
\begin{align}
&\rho_{+,ge}(\Delta_{\text{c}}) = 
\CR
& \frac{\ii \eta \tilde\Omega_{\text{p}} |\tilde\Omega_{\text{c}}|^{2} \tilde\omega_{\text{m}}}{(2\ii\tilde\Delta_{\text{c}}+|\tilde\Omega_{\text{c}}|^{2}) (2\ii [1-2\ii \tilde\omega_{\text{m}}] [\tilde\Delta_{\text{c}} -\tilde\omega_{\text{m}}] +|\tilde\Omega_{\text{c}}|^{2})},
\label{coherence_oscill}
\end{align}
where we have defined scaled quantities as $\tilde x = x/\Gamma_{\text p}$ and expanded $\hat{\rho}_{\pm}$ to first order in $\tilde\Omega_{\text{p}}$, requiring $\tilde\Omega_{\text{c}} \gg \tilde\Omega_{\text{p}}$ and $ \tilde\Omega_{\text{p}} \ll 1$, which amounts to typical EIT conditions. From \bref{coherence_oscill} we can determine ${\delta\rho}_{ge}''$ and $\alpha$ in \bref{coherencemodul} as $ {\delta\rho}_{ge}''=|\rho_{+,ge}(\Delta_{\text{c}}) - \rho_{+,ge}(-\Delta_{\text{c}})|$ and $\alpha = \arg[\rho_{+,ge}(\Delta_{\text{c}}) - \rho_{+,ge}(-\Delta_{\text{c}})] + \pi/2$, where $\mbox{arg}[z]$ is the argument of the complex number $z$. Here, we have used the expansion \bref{rhoexpand} 
and the fact that $\rho_{-,ge}^{*}(\Delta_{\text{c}})=\rho_{+,ge}(-\Delta_{\text{c}})$.

\begin{figure}[htb]
\centering
\includegraphics[width=1.0\columnwidth]{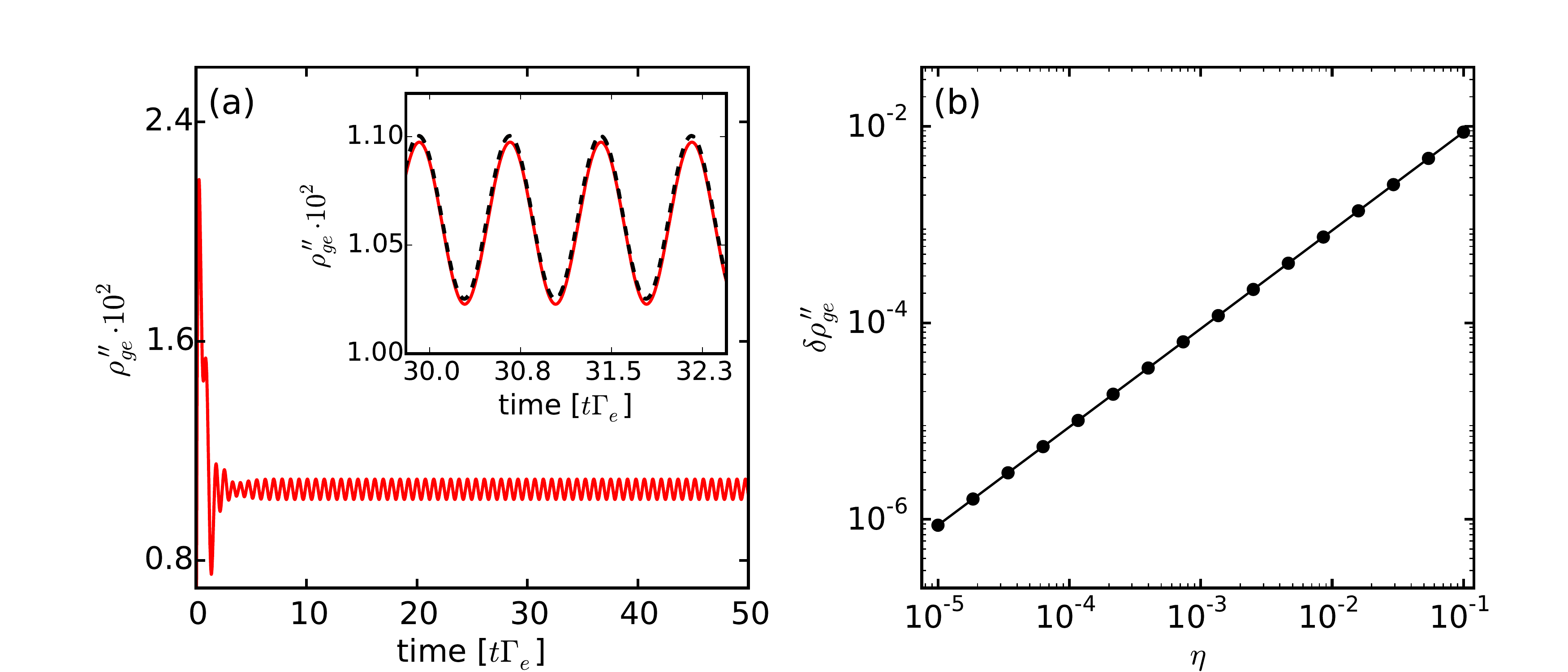}
\caption{(color online) Asymptotic response of atomic system to constant control beam sidebands, from numerical solutions of \eref{mastereqn}. We show the imaginary part of the probe transition coherence $\rho''_{ge}(t)$ that causes absorption, using $\Delta_{\text{c}} = 4.13 \, \text{MHz} \cdot 2\pi$, $\omega_{\text{m}} = 8 \, \text{MHz} \cdot 2\pi$, $\Omega_{\text{p}} = 0.32 \, \text{MHz} \cdot 2\pi$ and $\Omega_{\text{c}} = 10 \, \text{MHz} \cdot 2\pi$, $\Gamma_{\text{p}}= 6.1 \, \text{MHz} \cdot2\pi$. (a) Time evolution of $\rho''_{ge}$ for $\eta = 0.08$. The inset shows a zoom on the asymptotic behaviour, where the black dashed stems from our analytic solution \eref{coherence_oscill}. (b) The amplitude $\delta \rho_{ge}^{\prime \prime}$ of these oscillations scales linearly with the side-band strength $\eta$; ($\bullet$) are data points, and the line guides the eye. 
\label{constant_sidebands}}
\end{figure}

\frefp{constant_sidebands}{a} demonstrates that \eref{coherence_oscill} correctly describes the long-term evolution of the atomic system. We show $\rho_{ge}''(t)$ from a numerical solution to the master equation \bref{mastereqn} with Newton's equation \bref{newton}, ignoring 
the driving force in \eref{newton} ($F(t)=0$), but initialising mirror oscillations $z_{\text{m}}(t)=z_0 \cos{(\omega_{\text{m}} t)}$ with $z_0>0$.
 This numerical solution is compared with the predictions of \eref{coherencemodul} and \eref{coherence_oscill}. After an initial transient phase of the full model until $t\Gamma_{\text{p}}\lesssim 1$, the probe coherence is modulated at the mirror frequency with amplitude and phase described by \eref{coherence_oscill}. The modulation scales linearly with $\eta$, justifying our early truncation of the hierarchy resulting from \bref{steadyfourier}.
 Note that any mean coherence is nearly suppressed ($\rho_{0,ge}^{\prime \prime} \approx 0$).

Through changes in radiation pressure, the periodic modulation of the transparency of the atomic medium just discussed will give rise to a periodic driving of the mirror through \eref{probetransmisison}. This driving is automatically resonant. By determining the phase-relation between driving and mirror motion as well as the amplitude of this driving, we can predict the response of the mirror from classical mechanics. To this end we plot in \fref{periodic_response} ${\delta\rho}_{ge}''$ and $\alpha$ of \eref{coherencemodul} according to \eref{coherence_oscill} for various mirror frequencies $\omega_{\text{m}}$ and detunings $\Delta_{\text{c}}$.
The amplitude ${\delta\rho}_{ge}''$ is maximal approximately at $\Delta_{\text{c}} = \pm \Delta_{\text{max}}$, with
\begin{align}
\Delta_{\text{max}} = \frac{\omega_{\text{m}}}{2}\left[1 + |\tilde{\Omega}_{\text{c}}|^{2} - \sqrt{\big (1-|\tilde{\Omega}_{\text{c}}|^2 \big)^2 + \frac{|\tilde{\Omega}_{\text{c}}|^4}{\tilde{\omega}_{\text{m}}^2}} \right]
\label{maxmodul}
\end{align}
as shown in \frefp{periodic_response}{a} as red dashed line. \eref{maxmodul} is valid when $\Omega_{\text{p}}$ is small compared to other energies. We can further expand \eref{maxmodul} in the quantities $|\tilde\Omega_{\text c}|^{-2}$ and $\tilde\omega_{\text{m}}^{-1}$, which are small for cases considered here and get the even simpler expression
\begin{align}
\sub{\Delta}{max} \simeq \left|\frac{\Omega_{\cc}^2 -4 \omega_{\mm}^2}{4 \omega_{\mm}} \right|,
\label{maxmodul2}
\end{align}
which we will exploit in \sref{applicability}.

We can also see in \fref{periodic_response} that a wide range of relative phases between the amplitude modulation of the probe beam, and the phase modulation of the control beam (or mirror motion) can be accessed through variations of the detuning $\Delta_{\text{c}}$.
The physical origin of  the sharp features in \fref{periodic_response} is a resonance between the mirror frequency and energy gaps in the atomic system. To see this, let us decompose \bref{singleatomhamil} for one atom as $\hat{H}(t) = \hat{H}_0 + \hat{V}(t)$, where the perturbation is $\hat{V}(t)/\hbar = f(t) \hat{G}$, with $\hat{G} =  \frac{\ii}{2}\eta  [\Omega_{\text{c}} \hat{\sigma}_{es} - \text{h.~c.}]$ and $f(t)=\cos{(\omega_{\text{m}} t)}$. Let us define eigenstates $\ket{\varphi_j}$ of $\hat{H}_{0}$ via $\hat{H}_{0}\ket{\varphi_j} = E_j \ket{\varphi_j}$. We now assume the system has relaxed into the EIT ground-state, $\ket{\varphi_d}$, but otherwise we ignore spontaneous decay here. 

It is clear that whenever $\omega_{\text{m}}=|E_d-E_j|/\hbar$ for some $j\neq d$, the perturbation will cause resonant transitions to $\ket{\varphi_j}$. This is the case for $\omega_{\text{m}}$ fulfilling \bref{maxmodul2}. Since in this scenario the superposition of $\ket{\varphi_d}$ and $\ket{\varphi_j}$ will beat at the mirror frequency, also $\rho_{ge}$ is modulated with $\omega_{\text{m}}$. We have confirmed this picture using time-dependent perturbation theory.

\subsection{Interacting mirror and atomic cloud}
\label{fullycoupled}

Based on the previous section, we now determine the consequences of enabling feedback from the atomic medium onto the mirror through varying radiation pressure forces in \eref{newton} and \eref{radpressure}. We only consider radiation pressure from the modulated part of the probe beam, 
\begin{align}
F(t)=-W_{\text{p}0}A \,\text{Im}[\rho_{+,ge}\text{e}^{-\ii\omega_{\text{m}}t} +\rho_{-,ge}\text{e}^{\ii\omega_{\text{m}}t}] \text{,}
\label{driving_force}
\end{align}
thereby assuming that the mirror is already oscillating around a new equilibrium position 
\begin{align}
z_{\text{eq}}= 2 \frac{W_{\text{c}} + W_{\text{p}0}(1-A  \rho^{''}_{0,ge})}{M\omega_{\text{m}}^2c}\text{,} 
\label{newequil}
\end{align}
due to the radiation pressure by the control beam and the constant part of the probe beam. For simplicity we set $z_{\text{eq}}=0$ from now on.

\begin{figure}[htb]
\centering
\includegraphics[width=0.95\columnwidth]{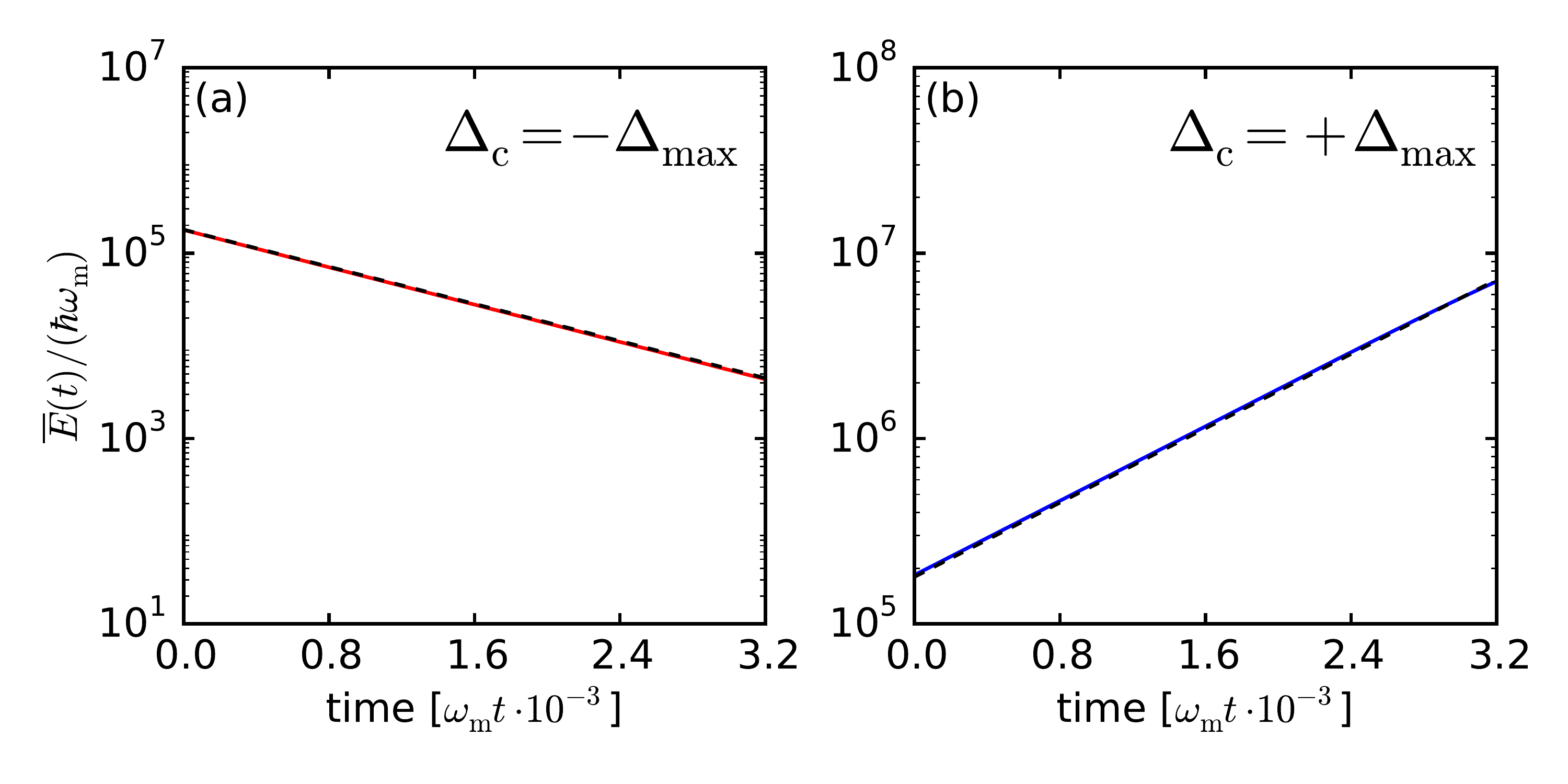}
\caption{Average energy $\bar{E}(t)$ of the vibrating mirror per mechanical period $T_{\text{m}}=2\pi/\omega_{\text{m}}$ in units of $\hbar \omega_{\text{m}}$ from a numerical integration of \eref{newton} and \eref{mastereqn}, using $M=10^{-20}$ kg, other parameters as in \fref{constant_sidebands}.
In this parameter regime, for red detuning $\Delta_{\text{c}} = -\Delta_{\text{max}}$, the mirror motion gets damped (a), whereas for a blue detuning $\Delta_{\text{c}} =+\Delta_{\text{max}}$ it gets amplified (b). Black dashed curves represent the model developed in \sref{fullycoupled} and we find good agreement. 
\label{coolandheat}}
\end{figure}

Using the driving force \bref{driving_force}, we numerically solve the coupled Newton equation \bref{newton} and master equation \bref{mastereqn}.
As can be seen in \fref{coolandheat}, the mirror can be driven such that its oscillation amplitude increases or decreases depending on $\Delta_{\text{c}}$. For a more quantitative description, we make the Ansatz $z_{\text{m}}(t) = Z(t)\cos{(\omega_{\text{m}} t)}$, where the amplitude $Z(t)$ is expected to vary very little during one mirror period $T_{\text{m}}$. The average energy of the oscillator per period is $\bar{E}(t)=1/2 M \omega_{\text{m}}^2 \bar{Z}^2(t)$. 
\begin{figure}[htb]
\centering
\includegraphics[width=1.0\columnwidth]{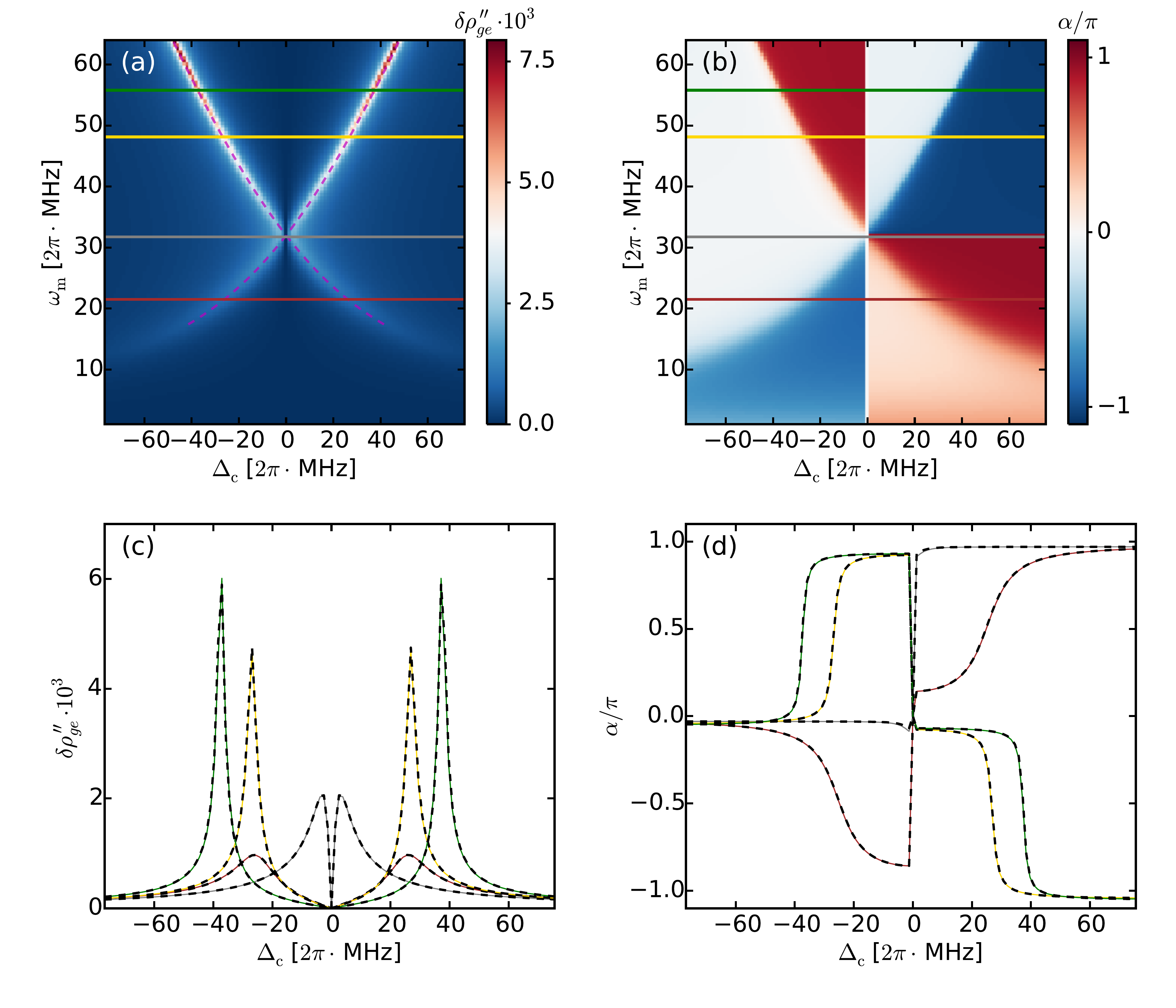} 
\caption{(color online) (a) Amplitude  $\delta \rho_{ge}^{\prime \prime}$ of the oscillations in the imaginary part of the probe transition coherence, $\rho_{ge}^{\prime \prime}$, as a function of mirror frequency $\omega_{\text{m}}$ and optical detuning $\Delta_{\text{c}}$, using \eref{coherence_oscill}.
 $\Omega_{\text{c}} = 64 \, \text{MHz} \cdot 2\pi$, $\eta=0.08$, other parameters as in \fref{constant_sidebands}. The red dashed line shows the peak position according to \eref{maxmodul}.
 (b) The phase $\alpha$ of the first harmonic of $\rho_{ge}^{\prime \prime}$ relative to the mirror oscillation, see \bref{coherencemodul}. Note that the apparent discontinuity for $\Delta_{\text{c}}>0$ is a meaningless $2\pi$ jump arising from plotting the phase continuously along the $\Delta_{\text{c}}$ axis. Plots in (c) and (d) are cuts through (a) and (b) respectively at the indicated values of mirror frequency $\omega_{\text{m}} = 21.3 \, \text{MHz} \cdot 2\pi$ (brown), $\omega_{\text{m}} = 32.0 \, \text{MHz} \cdot 2\pi$ (gray), $\omega_{\text{m}} = 48.0 \, \text{MHz} \cdot 2\pi$ (gold) and $\omega_{\text{m}} = 56.0 \, \text{MHz} \cdot 2\pi$ (green). Black dashed lines are a comparison of \eref{coherencemodul}-\eref{coherence_oscill} with direct numerical solutions of \eref{steadyfourier}.
\label{periodic_response}}
\end{figure}
Inserting the Ansatz into \eref{newton}, exploiting the slow variation of $Z(t)$ and using \eref{driving_force}, we find the solution 
\begin{align}
\bar{Z}(t) &= \bar{Z}(0) \text{e}^{-\Gamma_{\text{eff}}t/2}\text{,}
\label{effsolution}
\\
\Gamma_{\text{eff}} &=\frac{F_0 d \, \Gamma_{\text{p}}}{M\omega_{\text{m}} \Omega_{\text{p}}}\,\text{Re}[\rho_{+,ge}(\Delta_{\text{c}}) - \tensor*{\rho}{_{+,ge}}(-\Delta_{\text{c}})]
\CR
&=\frac{k_{\text{c}} F_0 d \, \Gamma_{\text{p}}}{M\omega_{\text{m}} \Omega_{\text{p}}}  \left[\frac{ {\delta \rho}_{ge}''}{\eta}\right]\sin(\alpha)\text{,}
\label{effdamprate}
\end{align}
where the overline denotes a time average over one mirror period. Details are shown in \aref{oscillator damping}. In \bref{effdamprate}, we use $F_0 = 2 W_{\text{p}0}/c$ and $\alpha$ can be determined from \bref{coherence_oscill}. Note that depending on the relative phase shift $\alpha$ between mirror motion and transparency modulations, the quantity $\sub{\Gamma}{eff}$ can actually describe damping or amplification. In \fref{periodic_response} we see that the effect on the mirror will be largest at the resonant feature near $\sub{\Delta}{max}$, with damping for negative detuning and amplification for positive detuning as long as $\omega_{\text{m}} < \Omega_{\text{c}}/2$. For $\omega_{\text{m}} > \Omega_{\text{c}}/2$ the two phenomena are swapped.

We validate the model \bref{effsolution} by comparing the predicted energy of a driven oscillator, using the analytical result for the atomic coherence \bref{coherence_oscill}, with the energy from a full numerical solution of Newton- \bref{newton} and master equation \bref{mastereqn}. We find good agreement as shown in \fref{coolandheat}. 

The results of the present section suggest an optical technique that makes use of atomic absorption to obtain a tailored optical driving force in order to control the mechanical state of a vibrating mirror.

\subsection{Range of applicability}
\label{applicability}

\begin{figure}[htb]
\centering
\includegraphics[width=1.0\columnwidth]{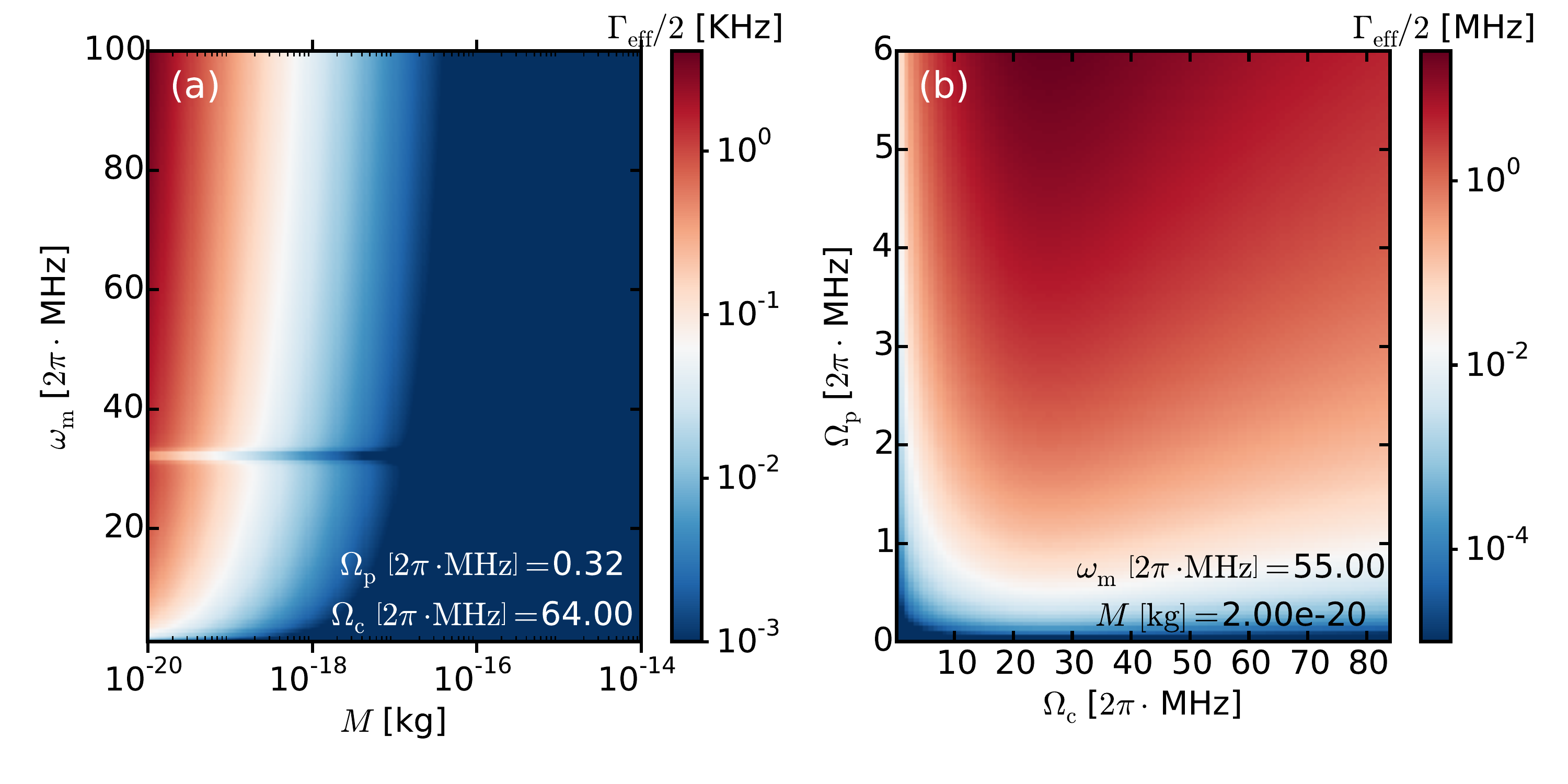}
\caption{Effective optical damping rate $\Gamma_{\text{eff}}$ for $\Delta_{\text{c}} = \Delta_{\text{max}}$. In (a) we show the dependence on the mirror mass $M$ and the mirror frequency $\omega_{\text{m}}$ for fixed Rabi frequencies $\Omega_{\text{p}}$ and $\Omega_{\text{c}}$; in (b) we fixed $M$ and $\omega_{\text{m}}$ and vary the Rabi frequencies.
\label{effective_damping}}
\end{figure}

For given oscillator parameters $\omega_{\text{m}}$ and $M$, the results of the preceding sections enable us to determine optical EIT parameters $\Omega_{\text{p},\text{c}}$, $\Delta_{\text{c}}$, for which the damping or amplification of the mirror is maximal (\eref{maxmodul}). For these we show the effective damping rate $\Gamma_{\text{eff}}$ of \bref{effdamprate} in \fref{effective_damping} for a variety of mirror parameters. 
Additionally, we also show the performance as a function of $\Omega_{\text{p},\text{c}}$ for fixed mirror parameters. 

Crucially underlying \fref{effective_damping} are our assumptions for light-field and medium properties. We have assumed a $^{87}$Rb medium of density ${\cal N} = 3.5 \cdot 10^{12} \, \text{cm}^{-3}$ and length $L = 242 \, \mu\text{m}$. Assigning the states $\ket{g}=\ket{5S_{1/2}, F = 1}$, $\ket{s}=\ket{5S_{1/2}, F = 2}$ and $\ket{e}=\ket{5P_{1/2}, F^{\prime} = 2}$, the Rabi-frequencies used in \fref{effective_damping} then roughly correspond to powers $W_{\text{p}0} \simeq 2.6 \cdot 10^{-2} \, \mu\text{W}$ and $W_{\text{c}} \simeq 3.2 \, \text{mW}$ at beam waists of $w_{\text{p}} = 350.0 \,\mu\text{m}$ for the probe- and  $w_{\text{c}} = 450.0 \, \mu\text{m}$ for the control beam. The transition frequencies used are $\omega_{\text{p}} \approx \omega_{\text{c}} \simeq 2.37 \cdot 10^{11} \, \text{MHz} \cdot 2\pi$. The decay rate $\Gamma_{\text{p}}\simeq 6.10 \, \text{MHz} \cdot 2\pi$. 
 
For this set of parameters, cooling rates in excess of typical environmental coupling strengths are accessible for rather light mirrors $M \lesssim 10^{-18} \text{kg}$ with frequencies $\omega_{\text{m}} \approx 10 - 100 \text{MHz} \cdot 2\pi$. Note that the feature at $\omega_{\text{m}}=\Omega_{\text{c}}/2$ is due to the absence of atomic response at this frequency, as evident in \frefp{periodic_response}{a}. Larger damping rates for heavier mirrors could be obtained with a larger probe power $W_{\text{p}0}$. In the present scheme, they are restricted by the requirement $\Omega_{\text{p}} < \Omega_{\text{c}}$. Variations of the scheme can be achieved by choosing a higher lying decaying state $\ket{e}$, which would decrease the transition matrix element $d_{eg}$ and thus allow larger $W_{\text{p}0}$ for identical Rabi frequency $\Omega_{\text{p}}$. However, simultaneously this would typically reduce the decay rate $\Gamma_{\text{p}}$.
%
\section{Conclusions and outlook}
\label{conclusion}

We have described an interface of the classical motion of a harmonically oscillating nano-mechanical mirror with the internal state dynamics and hence optical properties of a three-level, $\Lambda$-type atomic medium. Our atom-optomechanical setup exists in free-space, without any cavity. Coupling between mirror and atomic system is provided by the probe and control light fields that render the ultra cold atomic gas transparent, due to electro-magnetically induced transparency (EIT).

Depending on the choice of the EIT two-photon detuning, amplitude modulations of the probe light beam caused by the atomic medium 
are phase locked to the mirror oscillation. We have provided analytical expressions for the dependence of phase and strength of the modulations on the detuning.
The setup can also be seen as transferring phase modulations on one optical beam onto amplitude modulations of another.

When the modulated probe beam is made to interact with the mirror, oscillatory motion of the latter can be damped or amplified. We derive the effective damping (amplification) rate of the mirror, using a single atom type description of the EIT medium and a Fourier expansion of the density matrix in the presence of constant sidebands. The achievable damping rates exceed typical coupling strength of mirror to their thermal environment for light and fast mirrors ($M\lesssim 10^{-18}$ kg, $\omega_{\text{m}} \gtrsim 20 \, \text{MHz} \cdot 2\pi$). 

Our results provide the basis for a thorough understanding of the corresponding quantum-mechanical setup, which appears as a good candidate for a cavity-free cooling scheme \cite{camerer:latticecloud,vogell:longdistcouple}, that may complement established cavity cooling techniques \cite{aspelmeyer:review,kippenberg:review,marquard:coolingtheory,wildonrae:coolingtheory}. This will be the subject of future work.

Further interesting perspectives arise when our setup is extended towards Rydberg physics: EIT media where the second ground state $\ket{s}$ is replaced by a highly excited (and therefore also long-lived) Rydberg state $\ket{r}$ \cite{friedler:longrangepulses,Mohapatra:coherent_ryd_det,mauger:strontspec,Mohapatra:giantelectroopt,schempp:cpt,sevincli:quantuminterf,parigi:interactionnonlin}, have recently been used for the creation of single-photon sources \cite{dudin:singlephotsource,peyronel:quantnonlinopt} and proposed to enable nonlocal nonlinear optics \cite{sevincli:nonlocopt}. 
Much of the physics presented here is similar if $\ket{s}$ is replaced by a Rydberg state $\ket{r}$. Since this state $\ket{r}$ would be highly sensitive to interactions with other Rydberg atoms,
the control of mirror motion by further quantum mechanical atomic elements may be feasible also without an optical cavity. 

\acknowledgments

We gratefully acknowledge fruitful discussions with Yogesh Patil, Klemens Hammerer, Swati Singh, Igor Lesanovsky and Thomas Pohl, and EU financial support received from the Marie Curie Initial Training Network (ITN) COHERENCE".

\appendix

\section{Simplified optical response}
\label{optics}

For the system shown in \fref{setup}, the probe and control beams are coupled to an atomic medium with Rabi frequencies $\Omega_{\text{p}}$ and $\Omega_{\text{c}}$.
We assume the control field propagates inside the gas with group velocity $c$  and it is thus undisturbed by the response of the atoms \cite{GoAnLu07_033805}.
The evolution of the probe field is however determined by a wave equation in the presence of a source.
This source is the medium polarization at the probe field frequency, $P_{\text{p}} = P_{\text{p}}^{(+)} + \text{c.~c.}$,  given as the sum of its positive ($P_{\text{p}}^{(+)}$) and negative ($[P_{\text{p}}^{(+)}]^{*}$) frequency parts, respectively.
For a one dimensional description of the medium along $z$, the polarization is given by the collective slowly varying atomic coherence between $\ket{g}$ and $\ket{e}$, $\rho_{ge}(z,t)$, via $P_{\text{p}}^{(+)}(z,t)=d_{eg} \rho_{ge}(z,t) \exp[\ii (k_{\text{p}} z -\omega_{\text{p}}t)]$, where $\rho_{ij}(z,t)=\sum_n \langle \hat{\sigma}_{ij}^{(n)}(t)\rangle\delta(z - z_n)$.
Then within the slowly varying envelope approximation (SVEA) \cite{book:boyd_nonlinopt} the wave equation for the probe field reads
\begin{align}
\centering
[\partial_{t}+c \, \partial_{z}]\Omega_{\text{p}}(z,t) = \frac{\ii \omega_{\text{p}}}{2} \frac{ 6\pi \, \Gamma_{\text{p}}}{k_{\text{p}}^{3}} \rho_{ge}(z,t) \text{.}
\label{svea_eqns}
\end{align}
Eqs. \eref{svea_eqns} and \eref{mastereqn} form the so called set of Maxwell-Bloch equations.

Under usual stationary conditions of EIT one considers cw.~probe and control light fields impinging on the medium. 
It is then assumed that locally the density matrix elements $\rho_{ij}(z,t)$ settle into their steady state determined from $\dot{\hat{\rho}}=0$ in \eref{mastereqn}. 
Assuming a linear and homogeneous response of the medium we can define $\chi\Omega_{\text{p}} \equiv \frac{ 6\pi \, \Gamma_{\text{p}}}{k_{\text{p}}^{3}L}\rho_{ge}$.
The propagation \eref{svea_eqns} can now be analytically solved from $z=0$ to $z=L$ to yield 
\begin{align}
\centering
\Omega_{\text{p} L}&\approx  \Omega_{\text{p} 0}\left(1+ \ii k_{\text{p}}L \chi /2 \right) \text{,}
\label{Eout1}
\end{align}
under the condition $|k_{\text{p}} L \chi| \ll 1$.
Here $\Omega_{\text{p} 0}$ denotes the Rabi frequency of the incoming probe beam, while $\Omega_{\text{p} L}$ is that after passing through the medium of length $L$.

In our scenario the control beam has a residual time-dependence at the mirror frequency, as a result the probe beam is also modulated in time. As long as the propagation of the probe beam adiabatically follows the time evolution of the coupling beam, we can assume a modulated steady state to be locally attained everywhere in the medium, according to \eref{steadyfourier}. Considering the case in which retardation effects are negligible, $L \ll c/\omega_{\text{m}}$, we then integrate again \eref{svea_eqns} from $z=0$ to $z=L$ to
obtain a simple probe beam transmission through the medium as  
\begin{align}
\centering
\Omega_{\text{p} L}(t)&\approx  \Omega_{\text{p}0}\left(1+ \frac{\ii k_{\text{p}}}{2}L [\chi_0 + \chi_1 \text{e}^{-\ii \omega_{\text{m}}t} + \chi_{-1} \text{e}^{\ii \omega_{\text{m}}t}]\right)\text{.}
\label{Eout2}
\end{align}
For this, we assumed a linear response $\chi \Omega_{\text{p}0} = \frac{ 6\pi \, \Gamma_{\text{p}}}{k_{\text{p}}^{3}L} \rho_{ge}$, where $\chi$ is independent of $\Omega_{\text{p}0}$. This linearity was explicitly confirmed for cases considered here. 

\section{Effective damping of mirror's oscillation amplitude}
\label{oscillator damping}
%
The dynamics of the nano-mirror oscillations can be recast in terms of the complex variable $b_{\text{m}}(t) =  [z_{\text{m}}(t) + \ii \frac{p_{\text{m}}(t)}{M\omega_{\text{m}}}]$, with $p_{\text{m}}(t) = M\dot{z}_{\text{m}}(t)$ the canonical momentum associated to the displacement coordinate $z_{\text{m}}(t)$. 
 Newton's equation \bref{newton} is then equivalent to $\dot{b}_{\text{m}}(t) + \ii\omega_{\text{m}}b_{\text{m}}(t) = \ii \frac{F(t)}{M \omega_{\text{m}} }$, and damping and amplification of the nano-mirror motion will be reflected in the time evolution of its mechanical energy, $E(t) = 1/2 M \omega^2_{\text{m}} b^{*}_{\text{m}}(t)b_{\text{m}}(t)$.
Written in units of length and neglecting fast rotating terms ($\propto \text{e}^{\pm 2 \ii \omega_{\text{m}} t}$), the time evolution of the amplitude of motion of the mirror, $Z = \sqrt{b_{\text{m}}^{*}b_{\text{m}}}$, reads
\begin{align}
\dot{Z}(t)&\simeq \frac{F_0}{2M \omega_{\text{m}}}\frac{d \, \Gamma_{\text{p}}}{\Omega_{\text{p}}}\, \text{Re}[{\delta \rho}_{ge}''(t)\text{e}^{\ii (\alpha - \pi/2)}] \text{.}
\label{envelope_evolution}
\end{align}
Here ${\delta \rho}_{ge}''$, and $\alpha$ are derived in \sref{periodic} for constant mirror oscillations. Since the amplitude of mirror oscillations is now allowed to change in time, we make the replacement ${\delta \rho}''_{ge}\mapsto  [{\delta \rho}''_{ge}/\eta] k_{\text{c}} Z(t)$ in \eref{envelope_evolution}, where we used the linear dependence of ${\delta \rho}''_{ge}$ on the mirror oscillation amplitude $\sim\eta$ found in \sref{periodic}. The relative phase $\alpha$ does not depend on $Z(t)$ and hence remains constant.

We can finally solve equation \bref{envelope_evolution} coarse grained in time ($t > T_{\text{m}}$) by averaging over one mirror period to remove small variations of $Z(t)$, and obtain
\begin{align}
\dot{\bar{Z}}(t) = -\frac{k_{\text{c}} F_0}{2 M \omega_{\text{m}}}\frac{d \, \Gamma_{\text{p}}}{\Omega_{\text{p}}}\left[\frac{ {\delta \rho}_{ge}''}{\eta} \right]\bar{Z}(t)\sin{\alpha} \text{,}
\end{align}
with the solution \bref{effsolution}-\bref{effdamprate} in \sref{fullycoupled}.



\begin{thebibliography}{44}
\bibitem[{foo()}]{footnote:decay}
\bibinfo{note}{For simplicity we consider only decay from $\ket{e}$ to
  $\ket{g}$. In our parameter regimes, the physics is unchanged if decay from
  $\ket{e}$ to $\ket{s}$ is also possible.}
\end{thebibliography}
\end{document}